\documentclass[12pt]{article}

\def\lromn#1{\uppercase\expandafter{\romannumeral#1}}

\def\blist{\begin{list}{\setlength{\rightmargin}{\leftmargin}}}
\def\elist{\end{list}}
\begin{document}

\begin{flushright}
TU/00/601 \\
\end{flushright}

\begin{center}
\begin{large}

\textbf{
Resonance Enhanced Tunneling 
}

\end{large}

\vspace{36pt}

\begin{large}
Sh. Matsumoto$^{1}$ and M. Yoshimura$^{2}$

$^{1}$ 
Theory Group, KEK\\
Oho 1-1 Tsukuba Ibaraki 305-0801 Japan

$^{2}$ 
Department of Physics, Tohoku University\\
Sendai 980-8578 Japan\\
\end{large}

\vspace{4cm}

{\bf ABSTRACT}
\end{center}

Time evolution of tunneling in thermal medium is examined using
the real-time semiclassical formalism previously developed.
Effect of anharmonic terms in the potential well is shown to
give a new mechanism of resonance enhanced tunneling. 
If the friction from environment is small enough,
this mechanism may give a very large enhancement for the
tunneling rate.
The case of the asymmetric wine bottle potential is worked out
in detail.

\newpage

In our previous paper \cite{my 00-2} we formulated 
within the semiclassical framework how to deal with
the real-time dynamics of tunneling that takes place in thermal medium.
In the present work we further analyze the problem
and discover a new mechanism of enhanced tunneling
caused by anharmonic terms in the potential well.
The result may have great relevance to the old (once failed) scenario
of inflation based on the first order phase transition 
\cite{old inflation}, 
and to the scenario of electroweak baryogenesis \cite{ew-bgeneration-review}.

We start from recapitulating our previous result \cite{my 00-2}.
The real-time dynamics of tunneling in thermal medium is studied in
the standard model \cite{feynman-vernon},\cite{caldeira-leggett 83},
\cite{qbm review} of environment. Its interaction with
a one dimensional system which we describe by a potential $V(q)$
is of a bilinear form,
\( \:
q \int d\omega \,c(\omega )Q(\omega ) \,.
\: \)
Here $Q(\omega )$ is the environment oscillator coordinate of frequency
$\omega $ and $c(\omega )$ gives a coupling strength of
the system-environment interaction.
With the total system thus specified, dynamics is given by
the quantum equation of motion,
\begin{eqnarray}
&&
\frac{d^{2}q}{dt^{2}} + \frac{dV}{dq} =
-\,\int_{\omega_c}^{\infty}\,d\omega \,c(\omega )Q(\omega ) \,, 
\hspace{0.5cm}
\frac{d^{2}Q(\omega )}{dt^{2}} + \omega ^{2}\,Q(\omega )
= -\,c(\omega )\,q
 \,.
\label{eq of motion}
\end{eqnarray}

Quantum Langevin equation is derived \cite{ford-lewis-oconnell}
by eliminating the environment variable
$Q(\omega \,, t)$;
\begin{eqnarray}
&&
\frac{d^{2}q}{dt^{2}} + \frac{d V}{d q} +
2\,\int_{0}^{t}\,ds\,\alpha _{I}(t - s)q(s) = F_{Q}(t) 
\,,
\label{langevin eq}
\end{eqnarray}
where $F_{Q}(t) $ is linear in initial environment values,
$Q_{i}(\omega )$ and $P_{i}(\omega)$,
\begin{eqnarray}
&&
F_{Q}(t) = -\,
\int_{\omega _{c}}^{\infty }\,d\omega \,c(\omega )\,
\left( \,Q_{i}(\omega ) \cos (\omega t) +
\frac{P_{i}(\omega)}{\omega } \sin (\omega t)\,\right)
\,, 
\end{eqnarray}
and obeys the correlation formula,
\begin{eqnarray}
&&
\hspace*{-0.5cm}
\langle \{ F_{Q}(\tau )\,,F_{Q}(s) \}_{+} \rangle_{{\rm env}}
= \int_{\omega _{c}}^{\infty }\,d\omega \,r(\omega )
\cos \omega (\tau - s)\,\coth (\frac{\beta \omega }{2}) 
\equiv \alpha _{R}(\tau - s)
\,,
\end{eqnarray}
with $r(\omega ) = c^{2}(\omega )/(2\omega )$ and
$\beta = 1/T$ the inverse temperature.
The kernel function $\alpha _{I}$ in eq.(\ref{langevin eq}) 
is given by
\( \:
\alpha _{I}(t) = -\,\int_{\omega _{c}}^{\infty }\,d\omega \,
r(\omega )\sin (\omega t) 
 \,.
\: \)
The combination, $\alpha _{R}(t) + i\,\alpha _{I}(t)$,
is a sum of the real-time thermal Green's function, added with
the weight $c^2(\omega)$.

An often used simplification is the local, Ohmic approximation
taking 
\( \:
r(\omega) = \eta\omega/\pi
\: \)
with $\omega_c = 0$,
which amounts to
\( \:
\alpha _{I}(\tau ) = \delta \omega ^{2}\delta (\tau )
+ \eta\, \delta '(\tau ) \,.
\: \)
This gives the local version of Langevin equation,
\begin{equation}
\frac{d^{2}q}{dt^{2}} + \frac{d V}{d q} + 
\delta \omega ^{2}\,q + \eta\,\frac{dq}{dt} = 0 \,.
\end{equation}
The parameter $\delta \omega ^{2}$ is interpreted as
a potential renormalization or a mass renormalization in the
field theory analogy, since by changing the bare frequency parameter
to the renormalized $\omega _{R}^{2}$ the term
$\delta \omega ^{2}\,q$ is cancelled by the counter term
in the potential.
On the other hand, $\eta $ is the Ohmic friction coefficient.
This local approximation breaks down both at early and at late
times \cite{difficulty of ohm}, 
but it is useful in many other cases.

The crucial observation \cite{my 00-2} is
that in the semiclassical approximation the Wigner function 
has the solution in terms of classical solutions;
\begin{eqnarray}
&&
f_{W}(q\,, p\,, Q\,, P\,; t) = 
\int\,dq_{i}dp_{i}\,\int\,dQ_{i}dP_{i}\,
f_{W}^{(i)}(q_{i}\,, p_{i}\,, Q_{i}\,, P_{i})\,
\nonumber \\ &&
\hspace*{1cm} 
\cdot 
\delta \left( q - q_{{\rm cl}}\right)\,\delta \left( p - p_{{\rm cl}}\right)
\,\delta \left( Q - Q_{{\rm cl}}\right)\,\delta \left( P 
- P_{{\rm cl}}\right)
 \,.
\end{eqnarray}
Here 
\( \:
q_{{\rm cl}}(q_{i}\,, p_{i}\,, Q_{i}\,, P_{i}\, ; t)
\: \)
etc. are the solution of (\ref{eq of motion}), 
taken as the classical equation with the specified initial condition.
We take in the present work an uncorrelated initial state of the form
\( \:
\rho ^{(i)} = \rho _{q}^{(i)} \otimes \rho _{Q}^{(i)}
 \,.
\: \)
The reduced Wigner function is defined
after the environment variable ($Q(\omega) \,, P(\omega)$) integration;
\begin{eqnarray}
&&
f_{W}^{(R)}(q \,, p\,;t) = 
\int\,dq_{i}dp_{i}\,
f_{W\,, q}^{(i)}(q_{i}\,, p_{i})\,K(q \,, p\,, q_{i}\,, p_{i}\,;t)
\,,
\label{integral transform} 
\\ && 
K(q \,, p\,, q_{i}\,, p_{i}\,;t) = \int\,dQ_{i}dP_{i}\,
f_{W\,, Q}^{(i)}(Q_{i}\,, P_{i})\,
\delta \left( q - q_{{\rm cl}}\right)\,\delta \left( p - p_{{\rm cl}}\right)
 \,.
\label{def of kernel}
\end{eqnarray}

The next important point is expansion of the classical solution
$q_{{\rm cl}}$ in the environment initial variables $(Q_{i}\,, P_{i})$;
\begin{equation}
q_{{\rm cl}}(t) \approx  q_{{\rm cl}}^{(0)}(t) +
\int\,d\omega \,
\left\{\,Q_{i}(\omega )q_{{\rm cl}}^{(Q)}(\omega \,, t)
+ P_{i}(\omega )\,q_{{\rm cl}}^{(P)}(\omega \,, t)\,
\right\} \,.
\end{equation}
The truncation up to the first order term
in $(Q_{i}(\omega )\,, P_{i}(\omega ))$ 
should be adequate at low temperatures.
With the help of the familiar Fourier formula for the delta function
in eq.(\ref{def of kernel}),
one derives, using the Gaussian thermal function for the
initial $f_{W\,, Q}^{(i)}(Q_{i}\,, P_{i})$, 
an integral transform of the Wigner function, 
$f_{W\,, q}^{(i)}(q_{i}\,, p_{i}) \rightarrow f_{W}^{(R)}$. 
The resulting kernel function is given by \cite{my 00-2}
\begin{eqnarray}
&&
\hspace*{-1cm}
K(q \,, p\,, q_{i}\,, p_{i}\,;t) = 
\frac{\sqrt{{\rm det}\; {\cal J}\,}}{2\pi}
\exp \left[ \,-\,\frac{1}{2}
\left(q - q_{{\rm cl}}^{(0)}\,, \, p - p_{{\rm cl}}^{(0)} \right)
\,{\cal J}\,
\left( \begin{array}{c}
q - q_{{\rm cl}}^{(0)}  \\
p - p_{{\rm cl}}^{(0)}
\end{array}
\right)
\,\right]
 \,.
\label{kernel function} 
\end{eqnarray}
The matrix elements of 
\( \:
({\cal J}^{-1})_{ij} = I_{ij}
\: \)
are given as follows.
First,
\begin{eqnarray}
&&
I_{11} =\frac{1}{2}\,\int_{\omega _{c}}^{\infty }\,d\omega \,
\coth \frac{\beta \omega }{2}\,
\frac{1}{\omega } \,|z(\omega \,, t)|^2
\,, 
\label{width f}
\end{eqnarray}
and $I_{22}$ is given by a similar integral, replacing
$z(\omega \,, t)$ in $I_{11}$ by $\dot{z}(\omega \,, t)$, while
$I_{12} = \frac{\dot{I}_{11}}{2\,I_{11}}$.
Here
\( \:
z(\omega \,, t) = 
q_{{\rm cl}}^{(Q)}(\omega \,, t) + i\,
\omega \,q_{{\rm cl}}^{(P)}(\omega \,, t)\,,
\: \)
and 
\( \:
\dot{z}(\omega \,, t) = 
p_{{\rm cl}}^{(Q)}(\omega \,, t) +
i\,\omega \,p_{{\rm cl}}^{(P)}(\omega \,, t)
\,.
\: \)

Quantities that appear in the integral transform are determined by solving
differential equations; 
the homogeneous Langevin equation for $q_{{\rm cl}}^{(0)}$
and an inhomogeneous linear equation for $z(\omega \,, t)$ and
$\dot{z}(\omega \,, t)$,
\begin{eqnarray}
&&
\frac{d^{2}q_{{\rm cl}}^{(0)}}{dt^{2}} +
\left( \frac{d V}{dq }\right)_{q_{{\rm cl}}^{(0)}}
+ 2\,\int_{0}^{t}\,ds\,\alpha _{I}(t - s)q_{{\rm cl}}^{(0)}(s) = 0
 \,,
\label{homo classical eq}
\\ &&
\frac{d^{2}z(\omega \,, t)}{dt^{2}} +
\left( \frac{d^{2} V}{dq^{2} }\right)_{q_{{\rm cl}}^{(0)}}\,z(\omega \,, t)
+ 2\,\int_{0}^{t}\,ds\,\alpha _{I}(t - s)z(\omega \,, s) = 
-\,c(\omega )e^{i\omega t} \,.
\label{fluctuation eq}
\end{eqnarray}
A similar equation as for $z(\omega \,, t)$ holds for
\( \:
\dot{z}(\omega \,, t)  \,.
\: \)
The initial condition is \\
\( \:
q_{{\rm cl}}^{(0)}(t = 0) = q_{i} \,, \hspace{0.5cm} 
p_{{\rm cl}}^{(0)}(t = 0) = p_{i} \,, \hspace{0.5cm} 
z(\omega \,, t = 0) = 0  \,, \hspace{0.5cm}
\dot{z}(\omega \,, t = 0) = 0 \,.
\: \)

The physical picture underlying the formula for the integral
transform, eq.(\ref{integral transform}) along with 
(\ref{kernel function}), should be evident;
the probability at a phase space point $(q\,,p)$ is dominated 
by the semiclassical trajectory $q_{{\rm cl}}^{(0)}$
(environment effect of dissipation being included for its determination
by eq.(\ref{homo classical eq}))
reaching $(q\,,p)$ from an initial point $(q_i \,, p_i)$ 
whose contribution is weighed by the quantum mechanical probability 
$f^{(i)}_W$ initially given.
The contributing trajectory is broadened by the environment
interaction with the width factor $\sqrt{I_{ij}}$.
The quantity $I_{11}$ given by (\ref{width f}) and (\ref{fluctuation eq}),
for instance, is equal to
\( \:
\overline{(q - q_{{\rm cl}}^{(0)})^{2}}
\,;
\: \)
an environment driven fluctuation under the stochastic
force $F_{Q}(t)$.

The tunneling potential is divided into two regions separated 
at the position $q = q_B$ of the barrier top.
In the present work we shall assume that the potential is very steep
at both ends;
\( \:
V(\pm \infty ) = \infty \,.
\: \)
The tunneling rate, from the inner region at $q < q_B$
called here the potential
well into the outer region at $q > q_B$, is an important measure
of tunneling phenomena and is given by the flux at $q = q_B$;
\( \:
\dot{P}(t) = -\,I(q_{B} \,, t) \,,
\: \)
which is equal to
\begin{eqnarray}
&&
\hspace*{-1cm}
-\,
\int\,dq_{i}dp_{i}\,f_{W}^{(i)}(q_{i} \,, p_{i})
\,\left(\, 
p_{{\rm cl}}^{(0)} + \frac{\dot{I}_{11}}{2\,I_{11}}\,
(q_{B} - q_{{\rm cl}}^{(0)}\,)
\,\right)
\frac{1}{\sqrt{2\pi I_{11}}}\,\exp \left[ \,-\,
\frac{(q_{B} - q_{{\rm cl}}^{(0)}\,)^{2}}{2I_{11}}\,\right]
 \,.
\nonumber \\ &&
\end{eqnarray}
The width factor $\sqrt{I_{11}}$ thus determines the contributing
region of $q_{{\rm cl}}^{(0)}$ by how much away it is from $q_{B}$. 
On the other hand, the tunneling probability into the overbarrier
region at $q > x$ is given by
\begin{eqnarray}
&&
P(x \,, t) = 
\int\,dq_{i}dp_{i}\,f_{W}^{(i)}(q_{i} \,, p_{i})\,
\int_{x}^{\infty }\,du\,
\frac{1}{\sqrt{2\pi I_{11}}}\,\exp \left[ \,-\,
\frac{(u - q_{{\rm cl}}^{(0)}\,)^{2}}{2I_{11}}\,\right]
 \,.
\end{eqnarray}
In both of the quantities, $\dot{P}(t)$ and $P(q_B \,, t)$
the tunneling probability into $q > q_B$, 
it is essential to
estimate how $q_{{\rm cl}}^{(0)}$ and $I_{11}$ varies with time.

We shall assume an initial state localized in the potential
well so that the dominant contribution in the $(q_{i} \,, p_{i})$
phase space integration is restricted to $q_{i} < q_{B}$.
Let us first consider the harmonic
approximation for the potential well region given by
\( \:
V(q) + \frac{1}{2}\,
\delta \omega^2\,q^2 \approx \frac{1}{2}\,\omega_0 ^2 \, q^2
\: \)
near the bottom of the well at $q = 0$, and use the Ohmic
friction.
The approximate solution is given by
\begin{eqnarray}
&&
q_{{\rm cl}}^{(0)} = \left(\,\cos \tilde{\omega }_{0}t + \frac{\eta }{2
\tilde{\omega }_{0}}\sin \tilde{\omega }_{0}t \,\right)\,e^{-\eta t/2}\,q_{i}
 + \frac{\sin \tilde{\omega }_{0}t}{\tilde{\omega }_{0}}\,
e^{-\eta t/2}\,p_{i} \,, 
\\ && 
\hspace*{0.5cm} 
z(\omega \,, t) = 
\frac{c(\omega )}{\omega ^{2} - \omega _{0}^{2} - i\omega \eta }
\nonumber \\ &&
\cdot 
\left( \,
e^{i\omega t} - \frac{\omega + \tilde{\omega }_{0} - i\eta /2}
{2\tilde{\omega }_{0}}\,e^{i\tilde{\omega }_{0}t - \eta t/2}
+ \frac{\omega - \tilde{\omega }_{0} - i\eta /2}{2\tilde{\omega }_{0}}
\,e^{- i\tilde{\omega }_{0}t - \eta t/2}
\right)
 \,,
\end{eqnarray}
using
\( \:
\tilde{\omega }_{0} = \sqrt{\omega _{0}^{2} - \frac{\eta ^{2}}{4}}
\,.
\: \)
We assume a small friction,
\( \:
\eta \ll \omega _{0} \,.
\: \)

Near $\omega = \tilde{\omega }_{0}$ this function is approximately
\begin{eqnarray}
&&
z(\omega \,, t) \approx \frac{i\,c(\omega )}
{\omega + \tilde{\omega }_{0}- i\eta/2}
\,\left( \, t \,e^{i\tilde{\omega }_{0}t} - \frac{1}{\tilde{\omega }_{0}}\,
\sin \tilde{\omega }_{0}t\,\right)\,
e^{- \eta t/2}
 \,.
 \label{harmonic z}
\end{eqnarray}
This formula is valid at $t < O[1/\eta]$.
The appearance of the linear $t$ term is a resonance effect.
The resonance roughly contributes to $I_{11}(t)$ by the amount,
\( \:
\eta\, t^{2}\,  e^{-\eta t}\, \times
\, \)
a smooth $\omega $ integral which is cutoff by a physical 
frequency scale.
Thus, the width factor $I_{11}$ 
initially increases with time until the time
scale of order $1/\eta$.

At large times the classical trajectory $q_{{\rm cl}}^{(0)}$ asymptotically
approaches towards $q = 0$, the local minimum of the potential.
The width factor behaves as
\begin{eqnarray}
&&
I_{11}(t) = I_{11}(\infty ) + O[e^{-\,\eta t/2}]
\,,
\\ &&
I_{11}(\infty ) \approx  \frac{1}{2\omega _{0}} + \frac{1}{\omega _{0}}\,
\frac{1}{e^{\omega _{0}/T} - 1} + \frac{\pi }{3}
\frac{\eta }{\omega _{0}^{4}}\,T^{2}  \,.
\end{eqnarray}
The asymptotic value of $I_{11}(\infty )$ has
the familiar zero point fluctuation of harmonic oscillator 
and in the last term the dominant finite temperature correction,
valid for this Ohmic model at $T \ll \omega_0$.
In any event the probability rate $\dot{P}(t)$ finally decreases 
to zero, along with
\begin{equation}
p_{{\rm cl}}^{(0)} + \frac{\dot{I}_{11}}{2I_{11}}\,
(q_{B} - q_{{\rm cl}}^{(0)}\,) \rightarrow 0
\,. 
\end{equation}
Moreover, the final tunneling probability $P(q_{B} \,, \infty )$
has a finite value, and typically is very small for a large
potential barrier.
For instance, for the asymmetric wine bottle potential later
discussed,
\begin{equation}
P(q_{B} \,, \infty ) \approx \frac{1}{4}\,
\sqrt{\,\frac{\omega _{0}}{2\pi V_h}\,}
e^{-\,8 V_h/\omega_0} \,,
\end{equation}
with $V_h$ the barrier height much smaller than $\omega_0$.
This poses a curious question;
it appears that
decay of a prepared metastable state localized in the potential
well is never completed.

This simple picture is however valid only when one ignores anharmonic
terms in the tunneling potential, but they must be there
in order to give any realistic tunneling potential.
The most important point for the present work is effect of
anharmonic terms in the equation for the fluctuation $z(\omega \,, t)$.
Presence of anharmonic terms gives a non-trivial periodicity
in the coefficient function 
$\left(\frac{d^{2} V}{dq^{2} }\right)_{q_{{\rm cl}}^{(0)}}$
for (\ref{fluctuation eq}), assuming a small friction
$\eta \ll \omega_0$. For a very small friction the anharmonic term
becomes important both for completion
of the decay and the new mechanism of resonance enhanced tunneling.
In the rest of this paper we shall discuss the mechanism
of resonance enhancement.

One might naively expect that 
the homogeneous part of the $z(\omega \,, t)$ solution exhibits the well known
parametric resonance \cite{parametric resonance},
if the relevant parameter in the periodic coefficient function
$\left(\frac{d^{2} V}{dq^{2} }\right)_{q_{{\rm cl}}^{(0)}}$
falls in the instability band.
But this is not what happens, as will be shown.
Thus, unbounded exponential growth of $\sqrt{I_{11}}$ does not
take place.
On the other hand, the power-law growth is observed, as clearly seen
in numerical computation.
Moreover, we find that our $z(\omega \,, t)$ solution
belongs to the boundary between stability and instability bands.
At the resonance frequency the enhancement factor due to the boundary
effect is much larger than what one might expect from the harmonic
case (\ref{harmonic z}).
We shall interpret this phenomenon as influenced in a subtle
way by the parametric resonance, although it is not
the parametric resonance itself.

Observe first that if 
$q_{{\rm cl}}^{(0)}(t \,; \;q_{i} \,, p_{i})$ is a solution for
the classical, homogeneous Langevin equation, then
$q_{{\rm cl}}^{(0)}(t + \epsilon \,; \;q_{i} \,, p_{i})$
for an arbitrary constant time shift $\epsilon $ is also a solution.
Hence the time derivative
$\dot{q}_{{\rm cl}}^{(0)}(t \,; \;q_{i} \,, p_{i})
= p_{{\rm cl}}^{(0)}(t\,; \;q_{i} \,, p_{i})$
is a homogeneous solution for the linearized $z(\omega\,, t)$ equation.
Using this homogeneous solution, one can readily find the inhomogeneous
solution for $z(\omega \,, t)$ with the initial condition,
\( \:
z(\omega\,, 0) = \dot{z}(\omega\,, 0) = 0 \,.
\: \)
Corresponding to
the classical motion $q_{{\rm cl}}^{(0)}$ of given initial condition
$(q_i \,, p_i)$, this solution is
\begin{eqnarray}
&&
\hspace*{-1cm}
z(\omega \,, t) = -\,c(\omega )\,p_{{\rm cl}}^{(0)}(t)\,
\int_{0}^{t}\,dt'\,\left( p_{{\rm cl}}^{(0)}(t' )\right)^{-2}\,
\int_{0}^{t'}\,dt''\,e^{-\eta (t ' - t'')}\,p_{{\rm cl}}^{(0)}(t'' )\,
e^{i\omega t''}
 \,.
\label{inhomo z-sol}
\end{eqnarray}

It is best to discuss the resonance enhanced tunneling mechanism
in concrete examples. 
We take the asymmetric wine bottle potential as illustrated
in Fig.1, which is described in the well and its vicinity region by
\begin{equation}
V(q) \approx \frac{\lambda }{4}(q ^{2} - 2q_{B}q)^{2}
\,.
\end{equation}
The curvature parameters at two extema of $q=0$ and $q = q_B$ are 
\( \:
\omega _{0}^{2} = 2\lambda q_{B}^{2} \,, \hspace{0.5cm} 
\omega _{B}^{2} = \lambda q_{B}^{2} \,,
\: \)
and the barrier height seen from the bottom of the well is
\( \:
V_{h} = \frac{\lambda }{4}q_{B}^{4} = \omega _{0}^{2}q_{B}^{2}/8 \,.
\: \)
The condition for a large barrier, $V_{h} \gg \omega _{0}$,
implies 
\( \:
\sqrt{\lambda }\,q_{B}^{3} \gg 4\sqrt{2} \,.
\: \)
The classical $q_{{\rm cl}}^{(0)}$ equation in the Ohmic approximation
is written 
using rescaled variables, $y = q_{{\rm cl}}^{(0)}/q_{B}$ and
$\tau = \omega_0\,t/2$,
\begin{equation}
y'' + y(y - 1)(y - 2) + \frac{2\eta}{\omega_0}\, y' = 0 \,,
\label{classical ohmic langevin}
\end{equation}
where $y' = dy/d \tau$.

To give an idea of the magnitude of initial amplitudes, 
$y(0)$ and $y'(0)$,
let us assume an initial thermal state
trapped in the (hypothetical) harmonic
potential well, taking the same temperature as the environment.
The magnitude of initial values is then on the average
\( \:
\sqrt{\,\overline{q_{i}^{2}}\,} = \sqrt{\,
\coth \frac{\omega \omega _{0}}{2}/(2 \omega _{0})\,}
\approx \sqrt{1/(2\omega _{0})}
\,, \hspace{0.5cm} 
\sqrt{\,\overline{p_{i}^{2}}\,} \approx 
\sqrt{\omega _{0}/2}
\,.
\: \)
This gives
\begin{equation}
\frac{\sqrt{\,\overline{q_{i}^{2}}\,} }{q_{B}} =
\frac{\sqrt{\overline{p_{i}^{2}}}}{\omega _{0} q_{B}} =
\frac{1}{4}\,\sqrt{\frac{\omega _{0}}{V_{h}}} \,.
\end{equation}

The classical solution $p_{{\rm cl}}^{(0)}$ that appears in
the $z(\omega \,, t)$ solution (\ref{inhomo z-sol}) 
is given by the Jacobi's elliptic
function for the symmetric double well,
the situation relevant to the case of a small friction.
This function has the fundamental period given by
\begin{eqnarray}
&&
T = \frac{4}{\omega_0\,\sqrt{1 + \sqrt{\epsilon}}}\,
\int_{0}^{1}\,\frac{du}{\sqrt{(1 - u^{2})(1 - k^{2}u^{2})}}
\,, \hspace{0.5cm} 
k = \sqrt{\frac{2\sqrt{\epsilon }}{1 + \sqrt{\epsilon }} }
\,,
\end{eqnarray}
where
\( \:
\epsilon = E_i/V_h \,, \hspace{0.3cm}
E_i = p_i^2/2 + \omega_0^2\,q_i^2/2 \,.
\: \)
The corresponding fundamental frequency $\omega_*$ is $2\pi/T$.
The parameter $k$, hence $\epsilon$ is a measure of
the anharmonicity.
For instance,  $E(q_i \,, p_i) = V_h/4$ corresponds to $k \approx 0.80$,
a value giving $\omega_*$ different from the harmonic
frequency $\omega_0$ at the potential bottom; 
$\omega_* \approx 0.95 \times \omega_0$.
These formulas of the period are used to later determine
the band structure associated with the $z(\omega \,, t)$ equation.

We numerically integrated the coupled $q_{{\rm cl}}^{(0)}$ 
and $z(\omega \,, t)$ equations.
In the $\omega$ integral (\ref{width f}) for the width factor
$I_{11}$ the largest contribution is found to come from
the resonance at $\omega = \omega_*$, then contribution from
higher harmonics at $n\,\omega_*$ follows.
Thus, phenomenon of a non-linear resonance occurs.
In Fig.2 we show the width factor as a function of
time for a few choices of the initial data $(q_i \,, p_i)$.
For simplicity an equi-partition of the kinetic and the potential
energy 
\( \:
p_i^2/2 = \omega_0^2\,q_i^2/2
\: \)
is assumed to reduce relevant parameter dependence.
The result for the width factor is plotted for a few choices of
the initial energy.
At an intermediate time of $O[1/\eta]$ the width
factor becomes maximal at its value much larger than in
the harmonic case.
The decrease at late times is due to the friction;
we explicitly checked that for the zero friction, 
the fluctuation $|z(\omega \,, t)/c(\omega)|^2$ 
increases in time without a bound, with an averaged time power 
close to 4 at the resonance.
Note that even if the effect of the friction is turned off,
there exists an important environment effect here;
the environment interaction drives the non-linear resonance
oscillation.

The unbounded power-law increase suggests that the relevant parameter in
the homogeneous $z(\omega \,, t)$ equation falls in
the boundary between the stability and the instability bands,
since otherwise it either is bounded or grows exponentially.
We indeed checked this 
numerically by arbitrarily changing the parameters in
the $z(\omega \,, t)$ equation.
We used the homogeneous equation introducing
two arbitrary parameters $(h \,, \theta)$;
\begin{equation}
z'' + \left(\, h - 2\theta\,(\,y - \frac{1}{2}\,y^2 \,)\,\right)\,z = 0 
\,.
\label{eq for band}
\end{equation}
Unlike a single, linear equation for the Mathieu type
our problem is a coupled, non-linear system, with
$y$ to be determined by eq.(\ref{classical ohmic langevin})
for $\eta = 0$.
We must further assume some initial values for
$y(0) \,, \; y'(0)$.
The standard algorithm that determines the stability and the
instability band is used \cite{band cal}.
The parameter set $(h \,, \theta) = (4\,, 3)$ corresponds to
our homogeneous $z(\omega \,, t)$ equation for the zero friction.
This case is seen right on the boundary line of the two bands
as depicted in Fig.3.
We checked that our system is always on the boundary line
for all choices of $y(0) \,, \; y'(0)$ we computed for.

A rough analytic understanding of the growth of the width
factor seems to be as follows.
In the explicit formula for $z(\omega \,, t)$, eq.(\ref{inhomo z-sol}),
one may replace the quantity $\left(p_{{\rm cl}}^{(0)}(t' )\right)^{-2}$
in the integrand by some constant average value.
For a small friction it is then easy to obtain a linearly
growing $z(\omega \,, t)$ off the resonance and a quadratically
growing $z(\omega \,, t)$ right on the resonance as a function
of time in the range $t < O[1/\eta]$.
This time dependence is what is observed in numerical 
computation.

Computation of the tunneling rate $\dot{P}(t)$
is a demanding task of numerical calculation, and
it is left to future work.
But a general trend is expected already in the following simplified
computation.
The most important part for $\dot{P}(t)$ is
two competing exponential factors,
the one for the initial state and the other for the
kernel factor;
\begin{eqnarray}
&&
A = \exp \left[ \,- \,\,\tanh \frac{\beta \omega _{0}}{2}\,
\frac{p_i^2 + \omega_0^2\,q_i^2}{\omega _{0}}
\,\right]
\times
\exp \left[ \,-\,
\frac{(q_B - q_{{\rm cl}}^{(0)}\,)^{2}}{2I_{11}}\,\right]
 \,.
 \label{product factor}
\end{eqnarray}
We plotted in Fig.4 the time evolution of this product.
The time averaged evolution (averaged over a short time scale of
several times the fundamental period) shows a rapid rise
of the product factor,
reflecting the initial behavior, $I_{11} \propto t^4$.
The product then reaches some maximum at a time of order
$1/\eta$.
The asymptotic time limit coincides with that given by the
harmonic approximation.
As a numerical guide note that
\( \:
e^{-\,q_B^2/(2 I_{11})} = e^{-\,(2 V_h/\omega_0 )\,(\omega_0 I_{11}/2)^{-1}}
= e^{- \, 20\,(\omega_0 I_{11}/2)^{-1}}
\: \)
for $V_h = 10 \,\omega_0$.

In Fig.5 we show for a few choices of the friction
the maximum value of $A$ as a function of the initial energy $E_i$,
assuming the equi-partition $p_i = \omega_0 q_i$.
Remarkably, the largest contribution comes, not from the dominant
initial component near the zero point energy, 
rather from the initially suppressed excited component.
The first exponent in (\ref{product factor}) is in proportion to
$E_i$, while the second one goes roughly like $E_i^{-2}$,
hence a maximum may appear somewhere away from the lowest energy
of $E_i = \omega_0/2$.
In this example of $\eta/\omega_0 = 0.0025$ 
the maximum product factor is of order
$10^{-3}$ at $E_i \approx 4 \times \omega_0/2$, 
6 orders of magnitudes larger than
what one expects from the lowest energy state and also
the asymptotic value of order $10^{- 36}$.
A large value of the product factor of order unity 
suggests an interesting possibility of a rapid and violent termination
of the tunneling.

We end with some speculative comments on impact of
the resonance enhanced tunneling when applied to cosmology.
For this part of discussion we assume that the non-linear
resonance we found in the present work 
terminates the tunneling associated with
the asymmetric wine bottle type of potential.
The tunneling phenomenon is clearly related to termination
of the first order phase transition, if there exists a prolonged
period of cooling providing a chance that the universe, or
a part of it, is trapped in
a supercooled metastable state 
localized, say near $q = 0$ in our quantum mechanical terminology.
The coordinate $q$ here should be understood as the order parameter
of the phase transition, the homogeneous
Higgs field in the case of GUT \cite{old inflation}
and electroweak \cite{ew-bgeneration-review} phase transition.
The environment in this case is made of various forms of matter fields,
the quark and the lepton fields, and also 
the gauge and the  Higgs particle.
In the usual weak coupling theory the friction containing
typically small Yukawa couplings is small compared to
the typical energy scale of the system, the Higgs mass.
Thus the small dissipation seems a good approximation.
The low temperature approximation also works if
$T \ll $ the Higgs mass.
After many periods of the periodic Higgs motion 
(the period $\approx 2\pi/$(Higgs mass)\,) the resonance amplification
is expected to lead to an outburst of tunneling.
This may occur stochastically in different parts of the universe,
leaving behind nucleated bubbles, although
it is not clear how different bubbles subsequently merge.
Since the nucleation  occurs presumably within a short
time in all parts of the universe, it appears that
the first order phase transition is terminated abruptly.
This highly non-equilibrium incident should much help the
electroweak baryogenesis.

The other possibility of completing the phase transition
is signaled by the change of the potential  (more properly of the free energy)
due to the temperature change
so that the once local minimum at $q = 0$
becomes an unstable, local maximum of the potential.
Under a normal circumstance the rate of the temperature change
is given by the Hubble parameter $H$, hence if the Hubble parameter
is small enough, like $H < \eta$ 
(the friction), completion of the phase transition
via the resonance enhanced tunneling
most likely takes place. This condition seems to be satisfied in the
two cosmological examples given.

\vspace{0.5cm}
In summary, we gave using the real-time formalism of
tunneling dynamics 
a new mechanism of resonance enhanced tunneling.

\vspace{1cm}
\begin{center}
{\bf Acknowledgment}
\end{center}

The work of Sh. Matsumoto is partially
supported by the Japan Society of the Promotion of Science.

\vspace{1cm} 
%

\newpage
\begin{Large}
\begin{center}
{\bf Figure caption}
\end{center}
\end{Large}

\vspace{0.5cm} 
\hspace*{-0.5cm}
{\bf Fig.1}

A schematic form of asymmetric wine bottle potential.
A case of large barrier $V_h = 10 \,\omega_0 $ is depicted.
The barrier top is at $q_B = 4\sqrt{5}/\sqrt{\omega_0}$.

\vspace{0.5cm} 
\hspace*{-0.5cm}
{\bf Fig.2 } 

Time evolution of the width factor $I_{11}$.
A time average over $\omega_0 \Delta  t/2 = 20$ of
the dimensionless quantity $\omega_0 \,I_{11}/2$
is plotted as a function of time $\times \omega_0/2$.
Examples of initial states of energy 
$E_i = (1 \,, 2 \,, 5)\, \times \omega_0/2$
($\omega_0/2$ being the ground state energy in the potential well)
are compared to the harmonic case.
Assumed parameters are
\( \:
V_h = 10\,\omega_0 \,, \;
\eta/\omega_0 = 0.0025 \,, \;
T/\omega_0 = 1/20 \,, \;
\Omega ({\rm cutoff})/\omega_0 = 25
\,.
\: \)

\vspace{0.5cm} 
\hspace*{-0.5cm}
{\bf Fig.3 }

Band structure of stability (unshaded) and instability (shaded) for
the modified fluctuation equation.
The defining equation is (\ref{eq for band}), 
along with (\ref{classical ohmic langevin}), with the
initial condition of
\( \:
y(0) = y'(0) = 0.2
\,.
\: \)
The point marked by the cross has 
$(h \,, \theta) = (4 \,, 3)$ corresponding to
our system.

\vspace{0.5cm} 
\hspace*{-0.5cm}
{\bf Fig.4 }

Time averaged evolution of the
product $A$ of two exponential factors for the tunneling rate.
The same parameter set as in Fig.2 is taken.
In the inlet the case of $E_i =  5\, \times \omega_0/2$
is depicted in the linear scale.

\vspace{0.5cm} 
\hspace*{-0.5cm}
{\bf Fig.5 }

Maximal value of the product $A$
of two exponential factors for the tunneling rate is
plotted as a function of the initial energy for a few choices
of the friction $\eta$.
The other parameters are taken the same as in Fig.2.

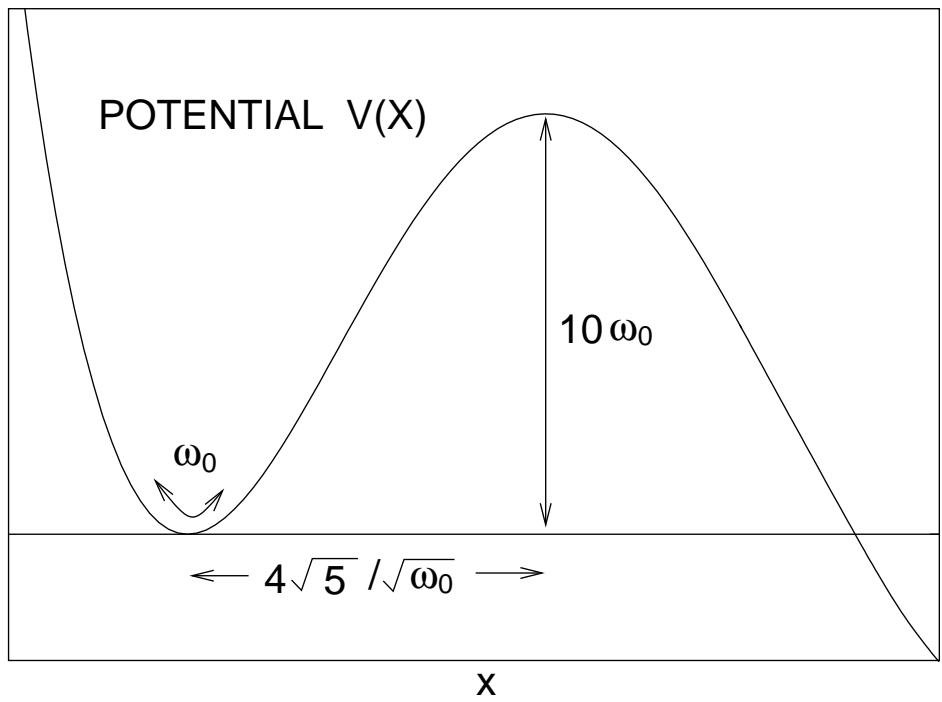
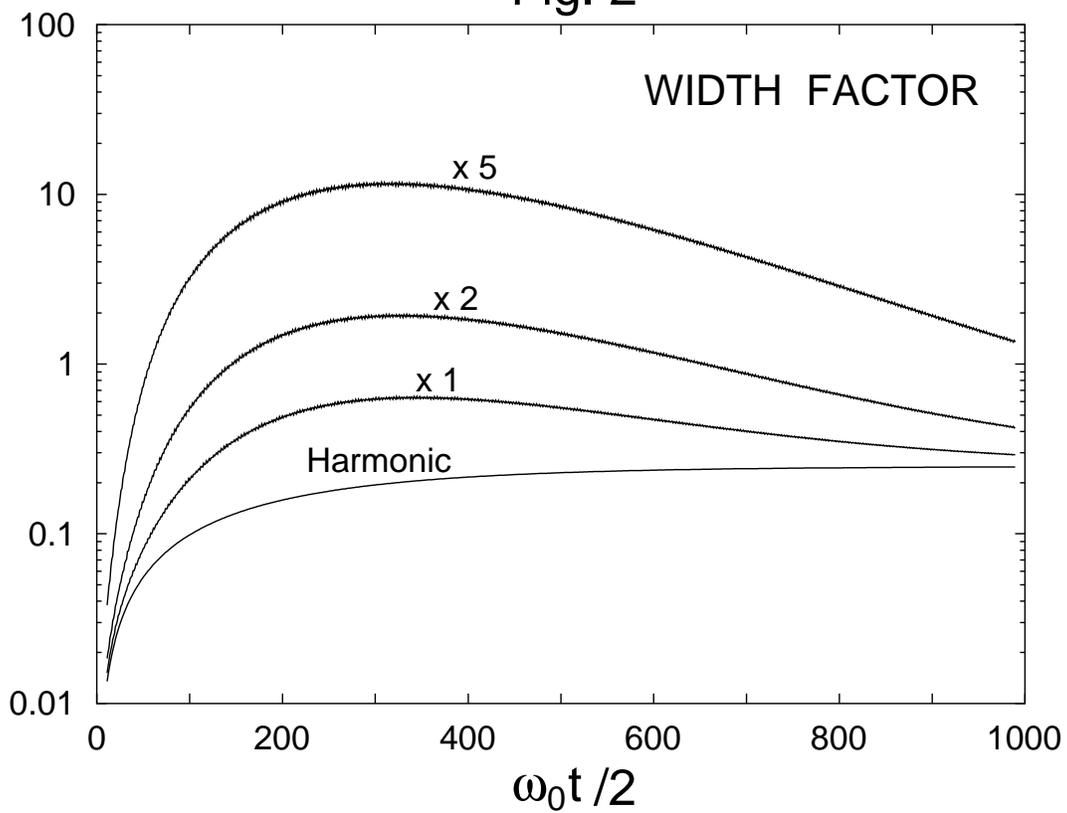
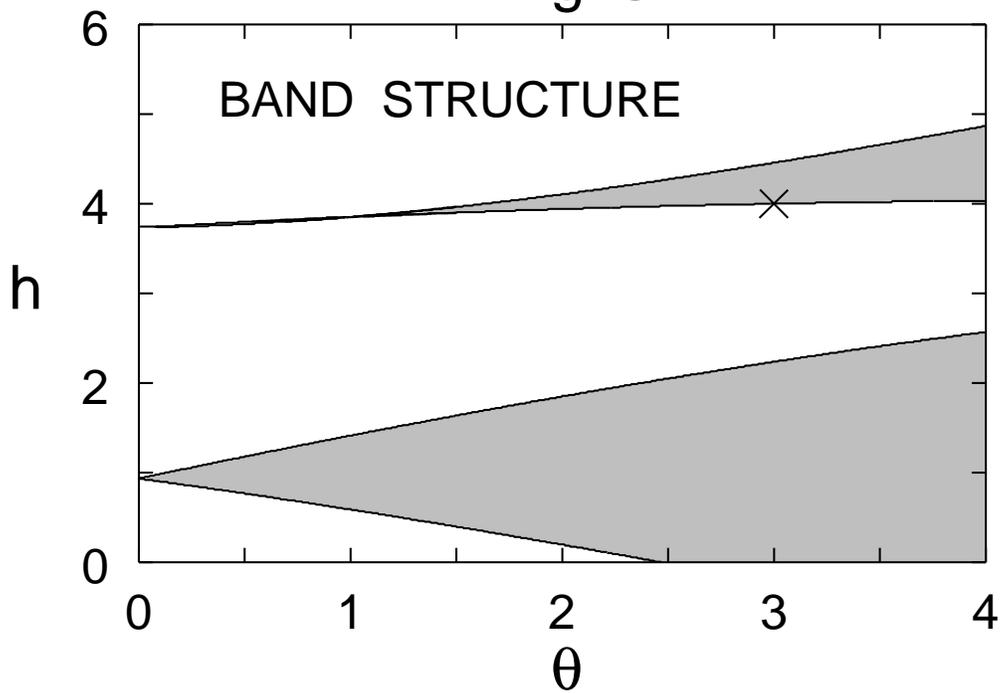
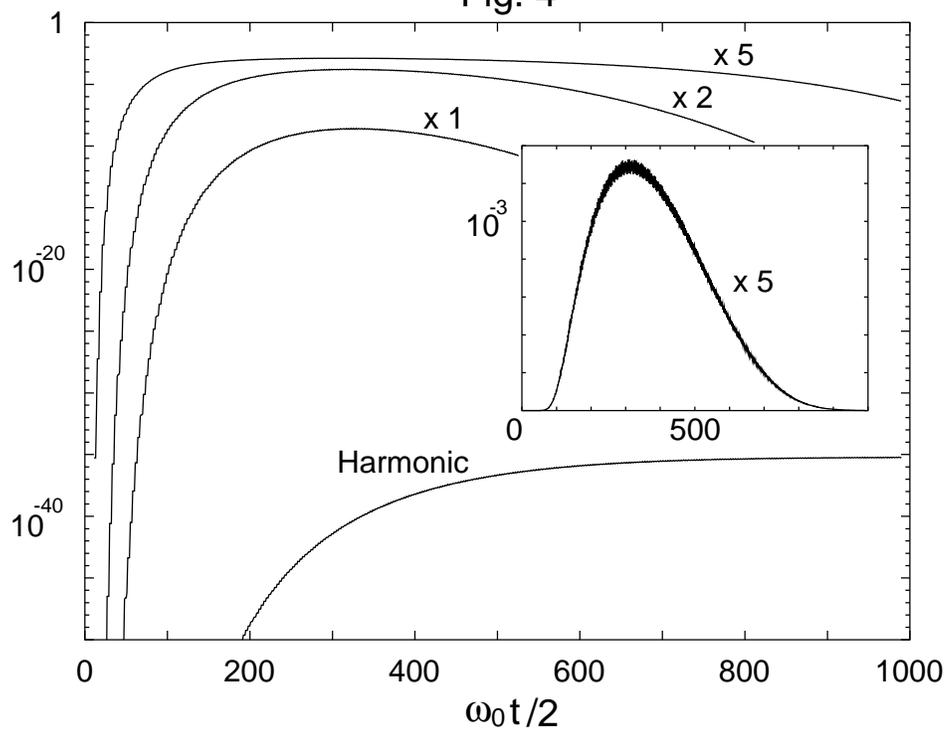
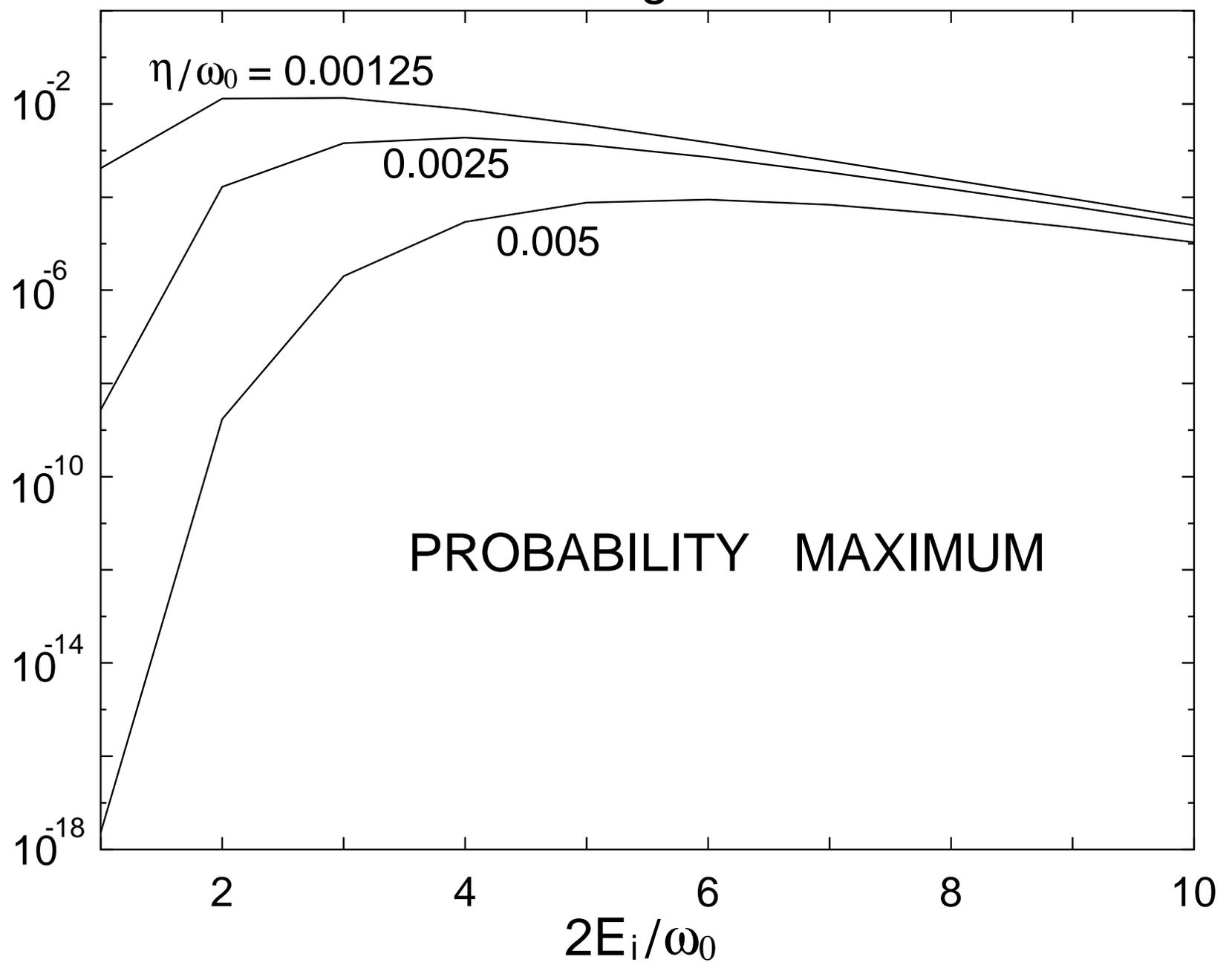
\end{document}